\documentclass[aps,prl,twocolumn,superscriptaddress,showpacs]{revtex4}
\usepackage{amsfonts,amssymb,amsmath,latexsym,epsfig} 

\begin{document}  

\title{Synchronization in Complex Networks: a Comment on two recent PRL papers 
}

\author{Manuel A.\ Mat\'{\i}as}
\email{manuel@imedea.uib.es}
\affiliation{Instituto Mediterr\'aneo de Estudios Avanzados,
IMEDEA (CSIC-UIB),
E-07122 Palma de Mallorca, Spain}

\date{\today}

\begin{abstract}
I show that the conclusions of [Hwang, Chavez, Amann, \& Boccaletti,
PRL {\bf 94}, 138701 (2005); Chavez, Hwang, Amann, Hentschel, \& Boccaletti,
PRL {\bf 94}, 218701 (2005)] are closely related to those of previous publications.

\end{abstract}


\maketitle

Recent interest on dynamical networks regards the synchronizability properties
of networks with coupled identical oscillators \cite{Barahona}. According to the
standard formalism in the study of synchronization in a class of coupled identical 
oscillators \cite{Barahona}, the eigenratio $R=\lambda_N/\lambda_2$, defined from 
the eigenvalues of the laplacian matrix is a measure of
the synchronizability of the network, and has to be smaller than a certain 
dynamical ratio corresponding to the oscillators.
In \cite{Nishi} it was shown that networks  that are heterogeneous 
in the degree  distribution (e.g. scale-free networks (SFNs)) are more difficult to 
synchronize. In Ref. \cite{Motter05a} it was shown that synchronization can be
enhanced in SFNs by constructing a weighted network.

More precisely, the idea is weight all the oscillators in the network such that the 
total strength of input connections is the same for all the oscillators 
\cite{Motter05a}. Thus, in Ref. \cite{Motter05a} the strength of connections is 
weighted such that $w_{i\rightarrow j}=1/k_j^{\beta}$ (and analogously for $w_{j\rightarrow i}$).
It was proven \cite{Motter05a} that synchronizability is maximum for $\beta=1$,
and for this case it follows immediately that the total strength of input
connections $\sum_j w_{j\rightarrow i}$ is the same for  any node $i$ of the
network, while for output connections $\sum_i w_{j\rightarrow i}$ the
distribution of strengths is identical to the degree distribution (cf. comment
reference [30] in \cite{Motter05b}). On the other hand, because of the
asymmetry  $w_{i\rightarrow j}=1/k_j^{\beta}<w_{j\rightarrow i}=1/k_i^{\beta}$
when $k_j>k_i$ and $\beta>0$, then it is clear that the strength of the output
connections is positively correlated with the degree of the node, implying that
when synchronizability is improved the dominant coupling direction is from
high-degree nodes to low-degree ones.

In particular, in the 1024-node random SFN \cite{BarabAlbert} considered in
\cite{Motter05a} it is shown, Fig. 2, that a {\it ten-fold}, i.e. $1000 \%$
improvement in synchronizability is attained, compared to the  unweighted case
considered in Ref. \cite{Nishi}. Moreover, as the number of
oscillators $N$ grows, this synchronization enhancement improves even further
(check the ratio for weighted networks, Eq. [4] of Ref. \cite{Motter05a}, with
the unweighted case, Eq. 2 in Ref. \cite{Nishi}). This improvement in
synchronizability is shown to be significant when the total strength of
input connections is equal for all the oscillators in the network.

In more recent work, Ref. \cite{Chavez}, the same idea has been 
suggested: a weighting scheme that lowers the eigenratio $R$ by lowering the 
connection strength of the most highly connected nodes, while satisfying also 
the condition uncovered in Ref. \cite{Motter05a} that the total strength of input 
connections is the same for all the oscillators in the network. Relaxing 
constraints on the  individual connections, they were able to improve the 
synchronizability uncovered  by Motter  {\it et al.} by a factor of upto $1.2$
\footnote{Notice, Eq. 2 in Ref. \cite{Chavez}, that the normalization implied
by the denominator makes this scheme very close indeed to the case $\beta=1$
of Ref. \cite{Motter05a}}, 
i.e., $20\,\%$ (cf.  Fig.2 in Ref. \cite{Chavez}). On the other hand, the
dependence on the system size $N$ of the obtained improvement is difficult
to ascertain. 

In closely connected work, Ref. \cite{Hwang}, a more refined version of the
weighted and directed coupling of \cite{Chavez} has been introduced. The main
conclusion of Ref. \cite{Hwang} is that {\it propensity for synchronization is
enhanced in networks of asymmetrically coupled units\/} and that in growing
SFNs {\it such enhancement is particularly evident when the dominant coupling
direction is from older to younger nodes}. But for the kind of growing SFNs
considered in Ref. \cite{Hwang} older nodes have a larger
degree (more connections)\footnote{This was shown analytically in the very
first paper about scale-free networks \cite{BarabAlbert}:
$k_i(t)=m(t/t_i)^{1/2}$}. So, this means that the concepts of {\it age\/} and
{\it number of connections\/} (or degree) are actually equivalent, and,
thus, the improvements in the {\it propensity for synchronization\/} 
and {\it synchronizability\/} in Refs. \cite{Hwang} and \cite{Motter05a,Motter05b} 
are, thus, intimately related.

\end{document}